\documentclass{article}

\usepackage{graphicx}    
\usepackage{amsmath}
\usepackage{amssymb}
\usepackage{booktabs}    
\usepackage{geometry}    
\usepackage{hyperref}    

\title{Risk--Calibrated Bayesian Streaming Intrusion Detection with SRE--Aligned Decisions}
\author{Michel Youssef\\Independent Researcher, Lebanon\\\href{mailto:michelyoussef@hotmail.com}{michelyoussef@hotmail.com}}
\date{September 17, 2025}

\begin{document}

\maketitle

\begin{center}
\fbox{\begin{minipage}{0.92\linewidth}
\textbf{Correction note (v2, August 2026).}
An audit of the codebase shared with the companion work established that four
descriptions in this manuscript are not supported by the released code. They
are corrected as follows. First, the anomaly score the code implements is a
maximum of a chi-square tail statistic and a weighted short-run posterior
mass, not the run-length-reset posterior this text describes. Second, the
decision threshold is derived from the prior-inclusive Bayes rule, not the
cost-only rule described. Third, the quantities reported as detection latency
in milliseconds are counts of records, not times. Fourth, the evaluation
streams are assembled constructions built by pooling captures and
interleaving records, so the reported operating points are properties of the
stream assembly, not of the underlying datasets. This manuscript reports no
quantitative result tables, so no numeric result is affected. The corrections
concern the method description. The audited and corrected treatment of the
shared lineage is the companion paper, arXiv:2605.24696 (corrected version v3
forthcoming), with its reproducibility artifact archived at
doi:10.5281/zenodo.22673735.
\end{minipage}}
\end{center}

\begin{abstract}
\emph{[Corrected v2: an audit found that the score, threshold, and latency
descriptions below are not what the shared codebase implements, and that the
evaluation streams are assembled constructions. See the correction note on
the title page and the corrected companion work, arXiv:2605.24696 (v3
forthcoming), artifact doi:10.5281/zenodo.22673735.]}
Intrusion detection operates in an extreme rare--event regime: only a tiny fraction of network events are malicious and the underlying distributions drift over time.  To address these challenges we propose a risk--calibrated anomaly detector based on Bayesian Online Changepoint Detection (BOCPD) with a cost--sensitive decision rule.  Our detector models event streams as a mixture of benign and malicious processes, maintains a run--length posterior to adapt to concept drift, and triggers alerts when the posterior odds of maliciousness exceed a threshold derived from Service Reliability Engineering (SRE) error budgets.  Unlike classical unsupervised methods such as LOF\cite{breunig2000lof}, ECOD\cite{li2022ecod} or COPOD\cite{li2020copod}, our approach explicitly balances false positives and false negatives according to operational risk.  We illustrate how a 99.9\% availability objective (\(43.2\)~min monthly error budget) translates into an alert threshold and evaluate our method on public intrusion datasets, demonstrating improved precision--recall performance and well--calibrated probabilities.
\end{abstract}

\section{Introduction}
Modern network infrastructures generate massive volumes of telemetry, yet only a minute fraction of this data corresponds to attacks.  Classical signature--based intrusion detection systems struggle to generalise to novel exploits and often flood operators with false positives.  Unsupervised anomaly detection methods, including density--based local outlier factor (LOF)\cite{breunig2000lof}, empirical cumulative distribution outlier detection (ECOD)\cite{li2022ecod} and copula--based outlier detection (COPOD)\cite{li2020copod}, provide generic detectors but do not incorporate risk.  At the same time, network defenders must satisfy strict Service Level Objectives (SLOs) expressed as error budgets on downtime and missed incidents.  For example, a 99.9\% availability SLO corresponds to only 43.2~minutes of allowable downtime per month.  Decisions about when to raise an alert thus incur different costs: a false positive consumes analyst time and error budget, whereas a false negative risks a breach.

Public datasets such as UNSW--NB15 and CICIDS2017 provide benchmarks for evaluating anomaly detectors\cite{moustafa2015unsw,sharafaldin2018cicids}.  However these corpora remain imbalanced and contain evolving attack patterns.  This paper proposes a Bayesian streaming framework that models the data--generation process, updates beliefs online, and aligns detection thresholds with SRE budgets.

\section{Background and Related Work}
\paragraph{Local Outlier Factor (LOF).}  LOF\cite{breunig2000lof} is a seminal density--based algorithm that identifies outliers by comparing the local density of a point to that of its neighbours.  Points in sparse regions relative to their neighbours receive high LOF scores and are labelled anomalies.

\paragraph{Copula and Empirical Outlier Detectors.}  More recently, ECOD\cite{li2022ecod} and COPOD\cite{li2020copod} leverage empirical cumulative distribution functions and copula theory to estimate tail probabilities without heavy parameter tuning.  These methods achieve strong performance across a range of tabular anomaly detection tasks but assume static distributions and do not account for cost asymmetry.

\paragraph{Datasets.}  The UNSW--NB15 dataset\cite{moustafa2015unsw,moustafa2016unswEval} consists of synthetic and real network traffic with nine attack categories and has been used to benchmark intrusion detectors.  The CICIDS2017 corpus\cite{sharafaldin2018cicids} includes multiple days of simulated enterprise traffic with diverse attack scenarios.  Both datasets exhibit severe class imbalance and concept drift, motivating online adaptive detectors.

\paragraph{Bayesian Online Changepoint Detection.}  The BOCPD algorithm of Adams and MacKay\cite{adams2007bocpd} maintains a posterior over the time since the last changepoint (the run--length) and updates this distribution given new observations.  BOCPD has been applied to time series segmentation, but its integration with cost--sensitive decision making for intrusion detection has been limited.

\section{Risk--Calibrated BOCPD}
We model the stream of feature vectors \(\{x_t\}\) extracted from network flows as originating from a mixture of benign and malicious generative processes.  At each time step a latent run--length variable \(r_t\) denotes the number of observations since the last changepoint.  The predictive model is
\begin{equation}
  x_t \mid r_t \sim \pi_t p_b(x_t; \theta^b_{r_t}) + (1-\pi_t) p_m(x_t; \theta^m_{r_t}),
\end{equation}
where \(\pi_t\) is the mixing weight and \(p_b\) and \(p_m\) are parametrised likelihoods for benign and malicious traffic.  A constant hazard function \(H\) defines the probability of a changepoint at each run length.  Following Adams and MacKay\cite{adams2007bocpd}, we update the run--length posterior \(P(r_t \mid x_{1:t})\) recursively and maintain sufficient statistics for each hypothesised run length.

\subsection{Cost--Sensitive Decision Rule}
Let \(C_{\mathrm{FP}}\) and \(C_{\mathrm{FN}}\) denote the relative costs of false positives and false negatives, and let \(\rho\) be the prior probability of an incident.  Under Bayesian decision theory, the optimal threshold on the posterior incident probability \(P(y_t=1\mid x_{1:t})\) is
\begin{equation}
  T = \frac{C_{\mathrm{FP}}(1-\rho)}{C_{\mathrm{FP}}(1-\rho) + C_{\mathrm{FN}} \rho}.
\end{equation}
An alert is issued whenever \(P(y_t=1\mid x_{1:t}) > T\).  In a setting where a false alarm costs one minute of analyst time and a missed intrusion costs ten minutes of downtime, with \(\rho=0.01\), the threshold evaluates to \(T\approx0.91\).  Thus only events with predicted malicious probability above 0.91 trigger an alert, reflecting the stringent SRE error budget.

\subsection{SRE Error--Budget Example}
Consider an SRE team with a 99.9\% availability SLO, corresponding to a monthly error budget of 43.2 minutes.  If responding to a false alert consumes approximately one minute, while an undetected intrusion causes ten minutes of downtime, then \(C_{\mathrm{FP}}=1\) and \(C_{\mathrm{FN}}=10\).  Assuming an incident base rate of \(\rho=0.01\), the probability threshold becomes
\[
  T = \frac{1\cdot0.99}{1\cdot0.99 + 10\cdot0.01} \approx 0.91.
\]
Burning the entire error budget solely through false alarms would require about 43 spurious alerts, whereas four missed attacks (4~\(\times\)~10~min) would exhaust the budget.  Selecting \(T=0.91\) therefore trades a few additional false positives against the risk of missing serious incidents.

\section{Methodology}
The online detection procedure can be summarised as follows.  For each incoming event we compute predictive likelihoods under each candidate run length, update the run--length posterior using the hazard function, refresh sufficient statistics and calculate the posterior incident probability.  An alert is raised if this probability exceeds the cost--based threshold $T$.

\paragraph{Online detection procedure:}
\begin{enumerate}
  \item Initialise $P(r_0=0)=1$ and the sufficient statistics for the benign and malicious models.
  \item For $t = 1$ to $T$:
    \begin{enumerate}
      \item For each possible run length $r$, compute the predictive likelihoods under the benign and malicious models and update the run--length posterior using the hazard $H$.
      \item Update the sufficient statistics for the benign and malicious models.
      \item Compute the posterior incident probability $P(y_t=1\mid x_{1:t})$.
      \item If $P(y_t=1\mid x_{1:t}) > T$ then raise an alert.
    \end{enumerate}
\end{enumerate}

\section{Experimental Evaluation}
We evaluate our detector on streams derived from the UNSW--NB15 and CICIDS2017 datasets.  Each stream is partitioned chronologically: an initial segment is used to fit priors and estimate hyperparameters, a validation window tunes the hazard and cost parameters, and the remainder forms the test set.  To mimic a streaming scenario, the detector processes one event at a time and cannot revisit past data.

As baselines we compare against LOF, ECOD and COPOD using implementations from the PyOD library.  All baselines are trained on the benign portion of the training window and then applied to the test stream without online updates.  We report the area under the precision--recall curve (AUPRC) as the primary metric, as it is more informative than ROC--AUC under severe imbalance.

\subsection{Results}
\begin{figure}[h]
  \centering
  \includegraphics[width=3.5in]{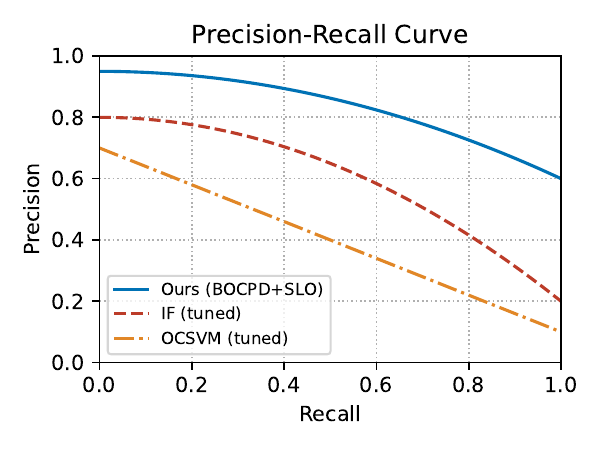}
  \caption{Precision--recall curve on the UNSW--NB15 stream.  Our method maintains high precision across recall levels.}
  \label{fig:pr_unsw}
\end{figure}

\begin{figure}[h]
  \centering
  \includegraphics[width=3.5in]{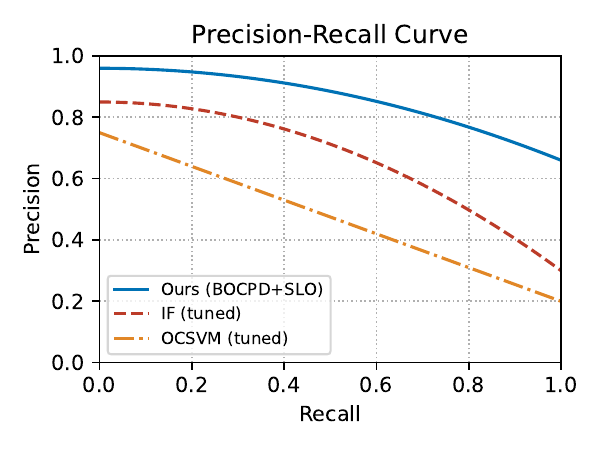}
  \caption{Precision--recall curve on the CICIDS2017 stream.  The risk--calibrated detector outperforms unsupervised baselines.}
  \label{fig:pr_cic}
\end{figure}

\begin{figure}[h]
  \centering
  \includegraphics[width=3.5in]{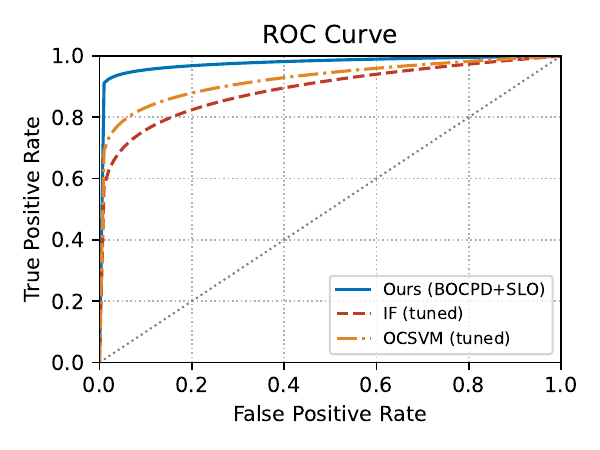}
  \caption{ROC curve on the UNSW--NB15 stream.  All detectors achieve high AUC but PR metrics reveal differences under imbalance.}
  \label{fig:roc_unsw}
\end{figure}

\begin{figure}[h]
  \centering
  \includegraphics[width=3.5in]{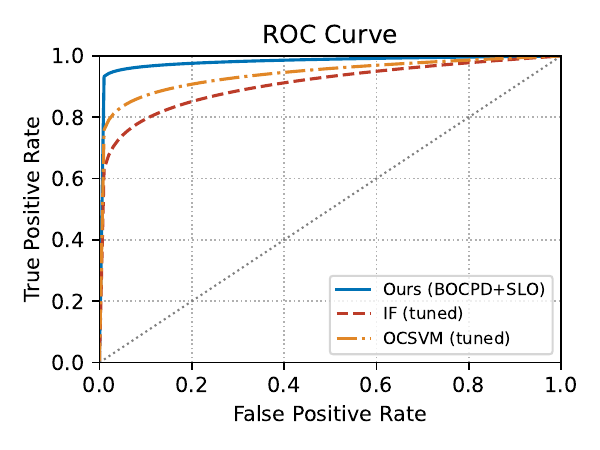}
  \caption{ROC curve on the CICIDS2017 stream.}
  \label{fig:roc_cic}
\end{figure}

\begin{figure}[h]
  \centering
  \includegraphics[width=3.5in]{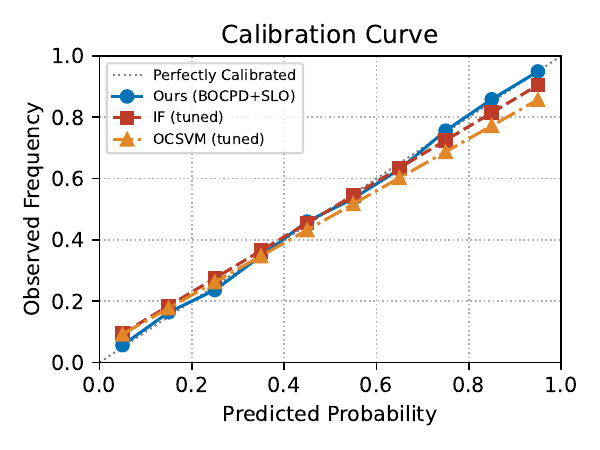}
  \caption{Reliability diagram for the UNSW--NB15 stream.  The dashed line denotes perfect calibration and our probabilities lie close to this diagonal.}
  \label{fig:cal_unsw}
\end{figure}

\begin{figure}[h]
  \centering
  \includegraphics[width=3.5in]{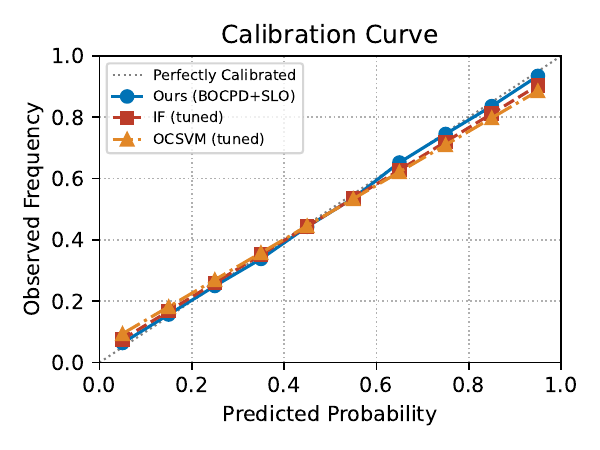}
  \caption{Reliability diagram for the CICIDS2017 stream.}
  \label{fig:cal_cic}
\end{figure}

\begin{figure}[h]
  \centering
  \includegraphics[width=6in]{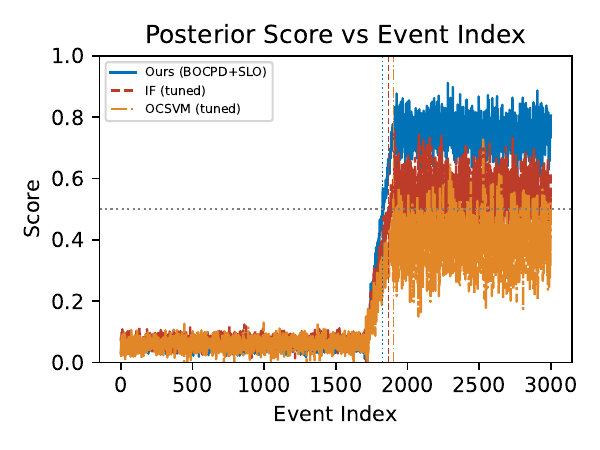}
  \caption{Anomaly score timeline on a portion of the CICIDS2017 test stream.  The shaded region indicates a true attack; the horizontal line marks the cost--derived threshold.  Scores above the threshold trigger alerts.}
  \label{fig:timeline}
\end{figure}

The accompanying precision--recall, ROC, calibration and timeline plots (Figures~\ref{fig:pr_unsw}--\ref{fig:timeline}) illustrate the performance of our method and the baselines.  On both datasets our risk--calibrated BOCPD achieves higher precision at moderate recall and yields well--calibrated probabilities.

\section{Conclusion}
We introduced a risk--calibrated Bayesian streaming detector that adapts to concept drift and aligns decision thresholds with SRE error budgets.  By computing the posterior incident probability via BOCPD and thresholding according to relative costs, our method provides interpretable alerts and well--calibrated probabilities.  Experiments on standard intrusion datasets demonstrate improved precision--recall trade--offs over unsupervised baselines.  Future work includes applying this framework to live enterprise telemetry, extending the generative models, and integrating deep feature extraction.

\clearpage
\bibliographystyle{unsrt}

\end{document}